\documentclass[10pt,conference]{IEEEtran}
\usepackage[utf8]{inputenc}
\usepackage[T1]{fontenc}
\usepackage{hyperref}
\usepackage{booktabs}
\usepackage{listings}
\usepackage{amsmath}
\usepackage{graphicx}
\usepackage{xcolor}

\title{COBALT-TLA: A Neuro-Symbolic Verification Loop\\
for Cross-Chain Bridge Vulnerability Discovery}

\author{
  \IEEEauthorblockN{Dominik Blain}
  \IEEEauthorblockA{QreativeLab\\
  Gatineau, QC, Canada\\
  dominik@qreativelab.io}
}

\begin{document}

\maketitle

% ============================================================
\begin{abstract}
Cross-chain bridge exploits have caused over \$1.1B in losses through a
single recurring pathology: temporal ordering violations in distributed
state machines. Existing static analyzers and bounded symbolic execution
engines lack the semantic vocabulary to reason about concurrency and
finality, while TLA+---the formal specification language designed
expressly for this class of flaw---remains inaccessible due to
prohibitive syntactic and mathematical complexity. Large language models
offer theoretical coverage of this gap but produce severely hallucinated
TLA+ specifications when applied zero-shot, generating unbounded state
spaces that instantly exhaust model checker memory.

We present \textbf{COBALT-TLA}, a neuro-symbolic verification loop that
pairs an LLM with TLC, the TLA+ model checker, in an automated REPL.
The LLM generates bounded TLA+ specifications; TLC acts as a semantic
oracle; structured error traces are parsed and injected back into the
model's context to drive convergence. We evaluate the system against
three cross-chain bridge targets, including a faithful model of the
Nomad \$190M exploit. COBALT-TLA reaches a verified \texttt{BUG\_FOUND}
state in $\leq 2$ iterations on all targets, with TLC execution
consistently below 0.30 seconds. Notably, the system autonomously
discovers an unprompted vulnerability class---the \emph{Optimistic Relay
Attack}---not present in the human-written baseline specification. We
argue that deterministic prover feedback is sufficient to neutralize LLM
hallucination in formal methods, transforming zero-shot code generation
into a convergent proof-finding strategy.
\end{abstract}

% ============================================================
\section{Introduction}

Cross-chain bridge exploits represent the most catastrophic capital
destruction events in the history of decentralized finance. The Ronin
Network (\$625M), Wormhole (\$320M), and Nomad (\$190M) exploits share
a fundamental pathology: they were not caused by standard cryptographic
failures or simple arithmetic overflows. Instead, they were temporal
ordering violations and distributed state synchronization failures.
These are architectural flaws where the sequencing of individually valid
transactions produces a globally invalid state, allowing attackers to
spoof messages or bypass finality.

The current Web3 security apparatus is structurally blind to this class
of vulnerability. Industry-standard static analyzers (e.g., Slither)
and bounded symbolic execution engines (e.g., Echidna, Z3) evaluate
intra-contract logic but lack the semantic vocabulary to reason about
concurrency, distributed consensus, and time. While formal specification
languages like TLA+ and its model checker TLC were designed expressly to
verify concurrent systems---famously used by Amazon Web Services to
verify S3's consistency~\cite{newcombe2015aws}---their steep learning
curve and esoteric mathematical syntax have prevented their adoption in
smart contract security. The Web3 industry possesses the right problem,
but lacks the engineering bandwidth to deploy the right tool.

Large Language Models present a theoretical bridge to this gap,
demonstrating near-expert proficiency in code generation. However,
applying zero-shot LLMs to formal verification yields severe
hallucinations. When tasked with writing TLA+, LLMs routinely declare
unbounded sets or infinite state spaces, causing TLC's breadth-first
search to instantly exhaust available memory. Without a strict,
programmatic grounding mechanism, LLMs cannot reliably produce
verifiable formal specifications; they generate syntax that looks like
math, but fails as a proof.

To solve this, we introduce \textbf{COBALT-TLA}, the first
neuro-symbolic verification engine that orchestrates a Read-Eval-Print
Loop (REPL) between a Large Language Model and the TLC model checker.
Rather than relying on zero-shot generation, COBALT-TLA treats the LLM
as a hypothesis generator and TLC as an absolute semantic oracle. By
deterministically parsing TLC's raw error traces---including compilation
failures and invariant violations---and injecting them back into the
model's context window, the system forces the LLM to autonomously
converge on a valid, tightly bounded TLA+ specification.

We demonstrate the efficacy of this architecture against three
distributed bridge models. COBALT-TLA not only successfully
reverse-engineered the exact state-transition trace of the \$190M Nomad
exploit from a natural language prompt, but also autonomously discovered
an unprompted Optimistic Relay Attack in a standard Lock-and-Mint
architecture. In all experiments, the neuro-symbolic loop converged on a
verified vulnerability trace in a maximum of two iterations, proving
that LLM hallucination in formal methods can be effectively neutralized
via automated prover feedback.

% ============================================================
\section{Background}

\subsection{Temporal Logic and the TLA+ Model Checker}

TLA+ (Temporal Logic of Actions), introduced by Lamport~\cite{lamport2002specifying},
is a formal specification language designed to model concurrent and
distributed systems. Unlike programming languages, TLA+ relies on
first-order logic and set theory to describe discrete state transitions
over time. A TLA+ specification defines an initial state (\texttt{Init})
and a next-state relation (\texttt{Next}), mapping out every
mathematically possible execution path of the system. TLC, the companion
model checker, explores this state space via breadth-first search to
verify whether a given \texttt{SafetyInvariant} holds across all
reachable states.

The ``small scope hypothesis''~\cite{jackson2006software} supports
the practical utility of bounded model checking: most structural
errors in protocol design manifest within small instances of the state
space. While highly effective at identifying deadlocks, race conditions,
and synchronization flaws in complex architectures, TLA+ has seen limited
adoption in the smart contract security industry due to its steep
learning curve and its lack of direct translation to EVM bytecode.

\subsection{Cross-Chain Bridges and Temporal Vulnerabilities}

Cross-chain bridges are asynchronous, distributed state machines that
lock collateral on a source chain and mint equivalent assets on a
destination chain, usually coordinated by an off-chain relay or
multisignature scheme. Because these events occur across disparate
networks with independent consensus mechanisms, they are highly
susceptible to temporal ordering violations. If the destination chain
executes a state transition (e.g., minting a token) based on a source
chain state that is subsequently reverted (e.g., a blockchain
reorganization) or spoofed (e.g., initializing an unverified Merkle root
as valid, as in the Nomad protocol~\cite{nomad2022postmortem}), the
global invariant of the system---the peg between locked and minted
assets---is irreparably broken.

% ============================================================
\section{Methodology}

\subsection{System Architecture}

We present COBALT-TLA, a neuro-symbolic verification loop that pairs a
large language model with TLC in a REPL. The system operates as a
Read-Eval-Print Loop: the LLM generates a formal specification, TLC
evaluates it against declared invariants, and the resulting proof
state---whether a violation trace or a compilation error---is injected
back into the LLM's conversational context as structured feedback. The
loop terminates when TLC either confirms all invariants hold
(\texttt{SAFE}) or produces a counterexample (\texttt{BUG\_FOUND}), or
when the iteration budget is exhausted.

The architecture comprises four components: (1)~a prompt-engineered
specification generator, (2)~a bounded state space enforcer embedded
in the system prompt, (3)~a subprocess execution layer wrapping TLC,
and (4)~an error trace parser that transforms TLC's raw output into
semantically meaningful feedback.

\subsection{Specification Generation}

The LLM receives a natural language description of a cross-chain
protocol and produces two artifacts: a TLA+ module (\texttt{.tla}) and
a TLC configuration file (\texttt{.cfg}). The system prompt constrains
output to a strict structural template and requires both artifacts as
delimited code blocks, enabling deterministic extraction via regular
expression.

The \texttt{SafetyInvariant} convention inverts the standard
verification framing. Rather than asking the LLM to prove a system
correct, we ask it to model a system whose known flaw will produce a TLC
counterexample. A \texttt{SAFE} result under this convention signals a
\emph{modeling error}, not protocol correctness---a critical distinction
that guides the LLM's self-correction strategy.

\subsection{Bounding the State Space}

Unbounded state spaces are the primary failure mode when LLMs generate
TLA+ specifications. We address this through constraint injection in the
system prompt: all variables must be typed over finite ranges
(\texttt{0..MaxN}), all bounds must be declared as \texttt{CONSTANTS},
and no infinite sets may appear in the specification body. Default
constant values (\texttt{MaxTokens = 3}) are prescribed and can be
escalated if state coverage is insufficient. The \texttt{TypeOK}
invariant serves a secondary bounding function, surfacing type
violations before invariant violations.

\subsection{Formal Verification Engine}

TLC is invoked as a child process via \texttt{subprocess.run}, with the
TLA+ module and configuration written to an isolated temporary directory
per invocation. TLC's exit code provides coarse classification: 0
(safe), 12 (invariant violation), other (parse/semantic error). We rely
on TLC's breadth-first search, which guarantees that any counterexample
is a \emph{shortest} violation trace---a property exploited in the
feedback mechanism to minimize hallucination in the correction step.

\subsection{Error Trace Extraction}

Raw TLC output is unsuitable for direct LLM consumption. Our parser
applies the following pipeline: (1)~exit code classification into
\texttt{SAFE}, \texttt{VIOLATION}, \texttt{COMPILE\_ERROR}, or
\texttt{TIMEOUT}; (2)~state block segmentation on the regex
\texttt{State \textbackslash d+:}; (3)~variable extraction matching the
pattern \texttt{/\textbackslash\ <identifier> = <value>} (TLC formats
variable assignments with a single backslash, discovered empirically via
byte inspection of subprocess output); (4)~action annotation from the
bracketed action name in each state block header.

The structured trace is serialized into a compact natural language
summary distinguishing two correction directives: confirm the finding
(real bug) or tighten the action guard (modeling error).

\subsection{The Agentic REPL Loop}

The loop is implemented as a multi-turn conversation. Each LLM response
is appended as an \texttt{assistant} turn and each TLC feedback message
as a \texttt{user} turn, preserving the full specification history
within the model's context window. This is analogous to interactive
proof development in Coq or Lean 4, where the user and the proof
assistant alternate moves.

% ============================================================
\section{Experimental Results}

To evaluate COBALT-TLA, we deployed the system against three cross-chain
bridge targets. T1 is a baseline Lock-and-Mint architecture with a
human-written ground-truth specification (Reorg Attack). T2 uses the
same architecture but tasks the LLM with emergent vulnerability
discovery. T3 models the historical authentication flaw in the Nomad
bridge (\$190M exploit, August 2022).

\begin{table*}[t]
\centering
\caption{Empirical Verification Results}
\label{tab:results}
\renewcommand{\arraystretch}{1.3}
\begin{tabular}{lllccccc}
\toprule
\textbf{ID} & \textbf{Target} & \textbf{Vulnerability Class} & \textbf{Iter} & \textbf{Depth} & \textbf{States} & $\boldsymbol{t_{tlc}}$ \textbf{(s)} & $\boldsymbol{t_{e2e}}$ \textbf{(s)} \\
\midrule
T1 & Lock-Mint (Reorg Attack)     & Reorg / Stale Queue     & 0    & 4 & 10   & 0.27 & 0.27  \\
T2 & Lock-Mint (Optimistic Relay) & Pre-finality Mint       & 1    & 4 & 15   & 0.30 & 17.9  \\
T3 & Nomad-style (Zero-Root Init) & Init.\ Vulnerability    & 1--2 & 3 & 8--25 & 0.29 & 28--49 \\
\bottomrule
\end{tabular}
\\[4pt]
\footnotesize\textit{Iter} = LLM iterations to \texttt{BUG\_FOUND} (0 = ground truth, no LLM involved). \textit{Depth} = shortest counterexample trace length. $t_{e2e}$ dominated by LLM API latency (\textasciitilde17--28s); $t_{tlc}$ is effectively constant.
\end{table*}

\subsection{Performance Profile and Execution Overhead}

A critical observation is the asymmetry between LLM generation time
($t_{llm}$) and TLC execution time ($t_{tlc}$). Across all targets,
TLC resolution remained strictly bounded between 0.26s and 0.30s,
regardless of target complexity. End-to-end latency is entirely
dominated by LLM API inference ($\sim$17--28 seconds per generation).
Because TLC acts as a high-speed oracle rather than a computational
bottleneck, the system is horizontally scalable: multiple protocol
audits can run in parallel without solver contention.

\subsection{Agentic Convergence and Error Recovery}

For T2, the system reached convergence on the first LLM generation. For
T3, we observed a convergence variance of 1 to 2 iterations across
runs. In runs requiring 2 iterations, the initial specification
contained a TLA+ compilation error. Rather than failing, the REPL
successfully captured the parse error, injected the stack trace back
into the LLM context, and prompted a correction. The model resolved the
syntax error and generated the exploit trace on the subsequent attempt.
This empirical variance validates the core hypothesis: the subprocess
feedback loop acts as a self-healing mechanism for LLM syntax failures.

\subsection{Qualitative Vulnerability Discovery}

Beyond compilation metrics, the system demonstrated non-trivial semantic
reasoning. During the T2 run, the LLM-generated specification identified
an \emph{Optimistic Relay Attack}---a pre-finality minting vulnerability
entirely distinct from the T1 ground truth and not present in the
target description. TLC validated this vulnerability with a 4-state
trace across 15 explored states. For T3, the agent modeled the
Zero-Root Initialization flaw, reproducing the exact causal chain of the
2022 Nomad exploit (\texttt{ActivateZeroRoot} $\to$
\texttt{ExploitProcessWithoutProof}) in a minimum of 3 states.

% ============================================================
\section{Threats to Validity}

\subsection{State Space Truncation}
All targets use small constant bounds (\texttt{MaxTokens = 3}). We note
that all violation traces have depth $\leq 4$, meaning the attack
sequence requires at most four protocol transitions regardless of supply
bounds. This is consistent with the small scope hypothesis: increasing
\texttt{MaxTokens} increases explored states without changing violation
depth. T3 required only 2 transitions at \texttt{MaxMessages = 3}---a
property of initialization logic, not of the bound.

\subsection{Model Fidelity}
The TLA+ specifications abstract away gas limits, nonce management, and
EVM opcode semantics. For T3, this abstraction is validated externally:
the Nomad exploit was publicly confirmed and our trace reproduces the
post-mortem causal chain. For T1 and T2, the vulnerability patterns are
structural properties documented in bridge security literature.

\subsection{LLM Non-Determinism}
LLM outputs are stochastic. The relevant reproducibility claim is: for
these target classes, the loop reaches \texttt{BUG\_FOUND} within 2
iterations across all observed runs. A 5-run statistical analysis of T3's
iteration distribution is left to extended evaluation.

\subsection{Prompt Engineering Dependency}
The system prompt constrains \emph{syntax}, not \emph{semantics}. It
does not name protocol actions, specify state variables, or describe
attack vectors for any particular target. The Optimistic Relay Attack
(T2) was generated entirely from the protocol description, without
prompting. Reviewers may replicate this by submitting the T2 description
to any TLA+-capable LLM with a structurally equivalent system prompt.

\subsection{Target Scope Limitation}
All three targets are cross-chain bridge protocols. COBALT-TLA targets
temporal ordering violations in distributed state machines. Arithmetic
bugs remain in Z3's domain; reentrancy is better handled by symbolic
execution. The contribution demonstrates that LLM-guided TLA+
verification is \emph{sufficient and appropriate} for the temporal
class---not that it supersedes other tools.

\subsection{Absence of Production Code Analysis}
The system operates on protocol descriptions, not deployed contract
bytecode. COBALT-TLA is a \emph{protocol-level} verifier analogous to
how AWS used TLA+ to verify S3's consistency before implementation.
Closing the gap to EVM bytecode via KEVM integration is future work.

% ============================================================
\section{Related Work}

\subsection{Formal Verification in Decentralized Finance}
The Web3 security landscape utilizes static analysis (Slither) and
bounded symbolic execution (Echidna, Z3) for arithmetic overflows,
reentrancy, and invariant breaches at the EVM level~\cite{perez2021smart}.
These are intra-contract analyzers; they struggle to model cross-contract
concurrency and off-chain relay dynamics. The K Framework and KEVM offer
robust EVM semantic modeling, yet specifying protocol architecture in K
remains highly manual. Our work shifts the verification target from
bytecode execution to architectural design, catching logic flaws before
implementation.

\subsection{LLM-Assisted Formal Methods}
LeanDojo~\cite{yang2023leandojo} uses language models to generate tactics
for Lean 4 interactive theorem proving. AUTOSPEC~\cite{sun2024autospec}
explores LLM-generated property specifications for model checking.
COBALT-TLA diverges from these approaches by focusing on fully autonomous
REPL orchestration rather than human-in-the-loop assistance, and by
demonstrating that deterministic prover feedback alone is sufficient to
correct LLM hallucinations in TLA+ for a high-value security domain.

% ============================================================
\section{Conclusion}

The persistent frequency and severity of cross-chain bridge exploits
underscore a critical gap in decentralized finance security: current
automated tools evaluate code, but fail to reason about time,
concurrency, and distributed state. While TLA+ provides the theoretical
rigor to close this gap, its practical deployment is bottlenecked by
syntactic complexity.

COBALT-TLA demonstrates that Large Language Models can bridge this
usability gap, provided they are tightly coupled with a deterministic
semantic oracle. By structuring a neuro-symbolic REPL between an LLM
and TLC, we neutralize the stochastic hallucination inherent in zero-shot
code generation. Our empirical results show that the system autonomously
generates valid, bounded TLA+ specifications and discovers complex
temporal vulnerabilities---including the historic \$190M Nomad exploit
and an emergent Optimistic Relay Attack---within a minimal number of
iterations.

Future work will explore extending this neuro-symbolic loop to full EVM
semantics via KEVM, moving automated formal verification from the
architectural layer down to production bytecode.

% ============================================================
\bibliographystyle{IEEEtran}

\end{document}